\documentclass[a4paper,11pt]{article}
\usepackage{aas_macros}
\usepackage{amsmath}
\usepackage{amsfonts}
\usepackage{amssymb}
\usepackage{graphicx}
\usepackage{jcappub}

\title{\boldmath Cosmic absorption of ultra high energy particles}

\author[a,b,c]{R. Ruffini}
\author[a,b,1]{G.V. Vereshchagin,\note{Corresponding author.}}
\author[a,b]{S.-S. Xue}

\affiliation[a]{ICRANet Piazzale della Repubblica, 10 -65122, Pescara}
\affiliation[b]{ICRA and Department of Physics, University of Rome ``Sapienza'', P.le A. Moro 5, 00185 Rome, Italy}
\affiliation[c]{ICRANet-Rio, Centro Brasileiro de Pesquisas Fisicas, Rua Dr. Xavier Sigaud 150, Rio de Janeiro, 22290–180, Brazil}

\emailAdd{ruffini@icra.it}
\emailAdd{veresh@icra.it}
\emailAdd{xue@icra.it}

\abstract{This paper summarizes the limits on propagation of ultra high energy particles
in the Universe, set up by their interactions with cosmic background of
photons and neutrinos. By taking into account cosmic evolution of these
backgrounds and considering appropriate interactions we derive the mean free
path for ultra high energy photons, protons and neutrinos. For photons the
relevant processes are the Breit-Wheeler process as well as the double pair
production process. For protons the relevant reactions are the photopion
production and the Bethe-Heitler process.\ We discuss the interplay between
the energy loss length and mean free path for the Bethe-Heitler process.
Neutrino opacity is determined by its scattering off the cosmic background
neutrino. We compute for the first time the high energy neutrino horizon as a
function of its energy.}

\begin{document}
\maketitle
\flushbottom

\section{Introduction}

Observation of ultra high energy (UHE) particles, such as photons, ions and
neutrinos, provides the crucial information on astrophysical systems as well
as mechanisms of charged particle acceleration in these systems. Such
information cannot be obtained from the study of low energy emission, although
it is much easier to detect.

Propagation of UHE particles on cosmological distances involves interaction
with other particles, as well as with electromagnetic fields, in the case of
charged particles \cite{Aharonian2003}. One of the most important reservoir of
photons is the cosmic microwave background (CMB). Interaction with
CMB\ imposes strong limits on propagation of UHE photons, protons, and nuclei.
Extragalactic background light (EBL), being the accumulated radiation in the
Universe\ due to stars and active galactic nuclei, represents additional
background of photons \cite{2001ARA&A..39..249H}, which limits propagation of
high energy photons. Yet another important background is cosmic neutrino
background (C$\nu$B), which places a tight limit on the propagation of UHE neutrinos.

In this work, we review stringent limits on propagation of UHE particles,
namely photons, protons and neutrinos, in the Universe due to their
interactions with cosmic background of photons and neutrinos. We pay the
particular attention to accounting for the cosmic evolution of CMB and C$\nu$B
fields, being important at high redshifts. We discuss relation with previous
results, as well as implications of new results, obtained in this work.

First in Sec. \ref{processes} we discuss most important processes, responsible
for interaction of UHE\ particles with cosmic backgrounds, as well as the
corresponding cross-sections. Most of these processes are discussed in detail
in \cite{2010PhR...487....1R}. Then in Sec. \ref{odandmfp} we introduce the
method used to compute the mean free path of UHE particles, which takes into
account cosmological redshift of particle energy, as well as temperature
evolution of the CMB and C$\nu$B. In Sec. \ref{meld} the definition of the
mean energy loss distance is given.\ We present and discuss results in Sec.
\ref{results}. Conclusions follow.

\section{Processes}

\label{processes}

\subsection{Processes involving photons}

UHE photons are likely produced in sources of UHE cosmic rays. The most
important process, responsible for intergalactic absorption of high-energy
$\gamma$-rays is the Breit-Wheeler process \cite{1934PhRv...46.1087B} for the
photon-photon pair production%
\begin{equation}
\gamma_{1}+\gamma_{2}\longrightarrow e^{+}+e^{-} \label{BWprocess}%
\end{equation}
It was first discussed by Nikishov \cite{Nikishov1961} back in 1961 and then,
after the discovery of CMB, by Gould and Schreder \cite{1967PhRv..155.1408G}.

Breit and Wheeler \cite{1934PhRv...46.1087B} studied collision process
(\ref{BWprocess}) of two photons with energies $E$ and $\mathcal{E}$ in the
laboratory frame, producing electron and positron pair. They found the total
cross-section
\begin{equation}
\sigma_{\gamma\gamma}=\frac{\pi}{2}\left(  \frac{\alpha\hbar}{m_{e}%
\,c}\right)  ^{2}(1-\beta^{2})\left[  2\beta(\beta^{2}-2)+(3-\beta^{4}%
)\ln\left(  \frac{1+\beta}{1-\beta}\right)  \right]  , \label{BW section0}%
\end{equation}
where
\begin{equation}
\beta=\sqrt{1-\frac{1}{x}},\quad x=\frac{E\mathcal{E}}{\left(  m_{e}%
c^{2}\right)  ^{2}}, \label{substitute}%
\end{equation}
$\hbar$\ is Planck's constant, $m_{e}$\ is electron mass, $c$\ is the speed of
light and $\alpha$\ is the fine structure constant. The necessary kinematic
condition in order for the process (\ref{BWprocess}) to take place is that the
energy of two colliding photons is larger than the energetic threshold
$2m_{e}c^{2}$, i.e., $x\geq1$. Due to this kinematic condition the function
(\ref{BW section0}) has a low energy cut-off at $x=1$. The cross-section has a
maximum at $x\simeq2$, with $\sigma_{\gamma\gamma}^{\max}\simeq\sigma_{T}/4$,
where $\sigma_{T}$ is Thomson cross-section. At higher energies it decreases
as $1/x$.

A simple estimate of the mean free path for the Breit-Wheeler absorption of
high energy photon can be given as follows. Considering the actual CMB photon
density $n_{CMB}\simeq411$ cm$^{-3}$ and taking $\sigma_{T}/4$ for the
cross-section of the interaction, the mean free path is $\lambda_{BW}=\left(
\sigma_{T}n_{CMB}/4\right)  ^{-1}\simeq4.8$ kpc. One can refer to this
distance as to a \emph{horizon}, namely the maximal distance to the source for
which the particle with the given energy can still be detected on Earth.
However, owing to the energy dependence of the cross-section, and cosmic
evolution of the CMB\ photon field the actual mean free path strongly depends
on energy. The characteristic energy of UHE photons interacting with CMB,
having temperature today $T_{0}\approx2.725$ K, is given by $E_{BW}%
=(m_{e}c^{2})/kT_{0}\simeq1.11$ PeV. At lower energies, in the TeV range,
photons interact by the Breit-Wheeler process with the EBL
\cite{1967PhRv..155.1408G,2000APh....12..217V,1999APh....11...35C}. Hence the
observation of TeV radiation from distant ($d>100$ Mpc) extragalactic objects
provides important constraints on the EBL
\cite{2007A&A...475L...9A,2012A&A...542A..59M,2014ApJ...795...91S}.

At much higher energies the double pair production process%
\begin{equation}
\gamma_{1}+\gamma_{2}\rightarrow e^{+}+e^{-}+e^{+}+e^{-} \label{dpp}%
\end{equation}
dominates, see \cite{1973ApL....14..203B,1997ApJ...487L...9C}. In this
high-energy limit it has nearly a constant cross-section, see e.g.
\cite{2010PhR...487....1R}
\begin{equation}
\sigma_{dpp}=\frac{\alpha^{2}}{36\pi}\left(  \frac{\alpha\hbar}{m_{e}%
\,c}\right)  ^{2}[175\zeta(3)-38]\sim6.5\mu b. \label{sigmaep2}%
\end{equation}
Clearly, this process has a threshold with the sum of energies of photons
which must exceed $4m_{e}c^{2}$. It imposes a limit for UHE photons
propagation $\lambda_{dpp}=\left(  \sigma_{dpp}n_{CMB}\right)  ^{-1}\simeq121$ Mpc.

\subsection{Processes involving protons}

Charged UHE particles, such as protons and nuclei, are assumed to originate
from extragalactic sources, which work as "cosmic accelerators"
\cite{Aharonian2003}. Such particles interact with the CMB\ photons as well.
In fact, the famous Greisen--Zatsepin--Kuzmin (GZK) limit
\cite{1966PhRvL..16..748G,1966JETPL...4...78Z} was established by considering
that this UHE particle interacts with the CMB\ photons via the pion
photoproducton process%
\begin{equation}
p+\gamma\longrightarrow\left(
\begin{array}
[c]{c}%
p\\
n
\end{array}
\right)  +\pi. \label{photopion}%
\end{equation}
and lose its initial energy. Due to the fact that at this process the proton
loses more than half of its energy \cite{2006NJPh....8..122D}, such
interaction imposes a strong cut-off on energies of UHE cosmic rays. The
cut-off energy is easy to estimate. Recall that the characteristic energy in
the Breit-Wheeler process (\ref{BWprocess}) is $E_{BW}=(m_{e}c^{2})^{2}%
/kT_{0}$. When the photopion process (\ref{photopion}) is concerned, the
electron mass is exchanged with the pion mass, and an additional factor 4
comes from the reference frame transformation, giving $E_{p\gamma}=4(m_{\pi
}c^{2})^{2}/kT_{0}\simeq3\times10^{5}E_{BW}=3.33\times10^{20}$ eV. More
careful evaluation of the energy by comparing energy losses due to photopion
(\ref{photopion}) and photoproduction of pair (\ref{ppair}) processes (see
below) gives the value $E_{p\gamma}=5\times10^{19}$ eV
\cite{1988A&A...199....1B}. The cross-section of the photopion process in high
energy limit is constant with the value \cite{2006NJPh....8..122D}%
\begin{equation}
\sigma_{p\gamma}\simeq120\mu b. \label{sigmapgamma}%
\end{equation}
The mean free path due to this process for energies $E>E_{p\gamma}$ is
$\lambda_{_{p\gamma}}=\left(  \sigma_{p\gamma}n_{CMB}\right)  ^{-1}\simeq6$ Mpc.

Another process relevant for interaction of UHE particles with CMB is the
photoproduction of electron-positron pair on a nucleus, or Bethe-Heitler
process \cite{1934RSPSA.146...83B}. In the case of proton, which is the only
one considered in this work, this process is%
\begin{equation}
p+\gamma\longrightarrow p+e^{+}+e^{-}. \label{ppair}%
\end{equation}
It has the characteristic energy $E_{BH}=m_{e}m_{p}c^{4}/(2kT_{0}%
)\simeq1.0\times10^{18}$ eV. This process has a threshold with photon energy
in the proton rest frame $\mathcal{E}^{\prime}>2m_{e}c^{2}$. We use for its
cross-section in the proton rest frame the expressions given in
\cite{1992ApJ...400..181C}, namely near the threshold with $2\leq
\epsilon^{\prime}\leq4$%
\begin{equation}
\sigma_{BH}^{thr}(\epsilon^{\prime})\simeq\frac{2\pi}{3}\alpha\left(
\frac{\alpha\hbar}{m_{e}\,c}\right)  ^{2}\left(  \frac{\epsilon^{\prime}%
-2}{\epsilon^{\prime}}\right)  ^{3}\left(  1+\frac{1}{2}\eta+\frac{23}{40}%
\eta^{2}+\frac{37}{120}\eta^{3}+\frac{61}{192}\eta^{4}\right)  ,
\label{sigmaphotopairthr}%
\end{equation}
where $\epsilon^{\prime}=\mathcal{E}^{\prime}/(m_{e}c^{2})$ is photon energy
in the proton rest frame and $\eta=\left(  \epsilon^{\prime}-2\right)
/\left(  \epsilon^{\prime}+2\right)  $. At higher energies $\epsilon^{\prime
}>4$ the cross-section is%
\begin{gather}
\sigma_{BH}^{he}(\epsilon^{\prime})\simeq\alpha\left(  \frac{\alpha\hbar
}{m_{e}\,c}\right)  ^{2}\left\{  \frac{28}{9}\delta-\frac{218}{27}+\left(
\frac{2}{\epsilon^{\prime}}\right)  ^{2}\left[  6\delta-\frac{7}{2}+\frac
{2}{3}\delta^{3}-\delta^{2}-\frac{\pi^{2}}{3}\delta+2\zeta(3)+\frac{\pi^{2}%
}{6}\right]  \right. \label{sigmaphotopair}\\
\left.  -\left(  \frac{2}{\epsilon^{\prime}}\right)  ^{4}\left[  \frac{3}%
{16}\delta+\frac{1}{8}\right]  -\left(  \frac{2}{\epsilon^{\prime}}\right)
^{6}\left[  \frac{29}{9\times256}\delta-\frac{77}{27\times512}\right]
\right\}  ,\nonumber
\end{gather}
where $\delta=\log(2\epsilon^{\prime})$. Expression (\ref{sigmaphotopair}) is
logarithmically increasing at high energies, so we can take a characteristic
value obtained by Bethe and Heitler $\sigma_{BH}\simeq(28/9)\alpha\left[
(\alpha\hbar)/(m_{e}\,c)\right]  ^{2}$ in order to estimate the mean free path
of UHE\ protons, which gives $\lambda_{BH}=\left(  \sigma_{BH}n_{CMB}\right)
^{-1}\simeq437$ kpc.

It is important to note that unlike the Breit-Wheeler process, leading to
annihilation of UHE photons, or the pion photoproducton, where single
interaction alters the energy of the UHE proton, the single Bethe-Heitler
interaction does not change the proton energy significantly. Therefore, unlike
all previous processes, the mean free path $\lambda_{BH}$ does not correspond
to a horizon. Another quantity is used for this purpose, namely the mean
energy loss distance, defined as $\lambda_{BH}\sim\lbrack dE/(Ecdt)]^{-1}$,
where $E$\ is the proton energy, which corresponds to the distance on which
the energy of the UHE proton is reduced by a factor $e$ due to numerous
interactions with background photons
\cite{2006NJPh....8..122D,1988A&A...199....1B,2000PhRvD..62i3005S}. However,
it should be emphasized that single Bethe-Heitler interaction deflects the UHE
proton by a small angle. This effect is discussed in detail below.

\subsection{Processes involving neutrinos}

UHE\ neutrinos can be produced either in astrophysical sources, or in some
exotic new physics scenarios \cite{2006JPhCS..39..393R}. Below we compute the
horizon due to interaction of UHE\ neutrinos with cosmic neutrino
background (C$\nu$B). Following \cite{2013JCAP...08..014L} we assume C$\nu$B
neutrinos are in their mass states.\ The cross-section is composed of two
parts \cite{2013JCAP...08..014L}. The resonant neutrino annihilation occurs in
the s-channel:%
\begin{equation}
\nu+\bar{\nu}\longrightarrow Z^{0}\longrightarrow f+\bar{f}, \label{Rnunubar}%
\end{equation}
where bar denotes antiparticle, $f$ is a fermion. It has a typical
Breit-Wigner shape and is given in the analytic form in
\cite{2006APh....25...47D}. We take the small momentum expansion of the
cross-section given by eq. (23) in \cite{2006APh....25...47D}\ as%
\begin{equation}
\sigma_{\nu\bar{\nu}}^{R}\simeq4\sqrt{2}G_{F}\frac{m_{\nu}M_{Z}^{2}\sqrt{\xi
}E}{(M_{Z}^{2}-2Em_{\nu})^{2}+4E^{2}m_{\nu}^{2}\xi}\text{GeV}^{-2},
\label{SigmaRnunubar}%
\end{equation}
where $G_{F}=1.16637\times10^{-5}$ GeV$^{-2}$ is the Fermi's coupling
constant, $\xi=\left(  \Gamma/M_{Z}\right)  ^{2}$, $\Gamma=2.495$ GeV is the
width of $Z^{0}$\ resonance and $M_{Z}=91.1876$ GeV is the mass of $Z^{0}$
boson, $m_{\nu}$\ is neutrino mass, $E$\ is energy of UHE\ neutrino in
laboratory frame. Clearly, the position of the resonance scales inversely
proportional to the neutrino mass. Throughout this paper we use the reference
value $m_{\nu}=0.08$ eV/c$^{2}$, corresponding to the recent cosmological
bound from the Planck mission \cite{2014A&A...571A..16P}, which gives
characteristic energy $E_{r}=M_{Z}^{2}c^{2}/2m_{\nu}\simeq5.2\times10^{22}$
eV. The amplitude of the resonance does not depend on neutrino mass, and is
given by
\begin{equation}
\sigma_{\nu\bar{\nu}}^{R\max}=2\sqrt{2}G_{F}M_{Z}/\Gamma\simeq0.471\mu b.
\label{SigmaRmax}%
\end{equation}

The second contribution is the non-resonant cross-section, which is adopted
here in the form%
\begin{equation}
\sigma_{\nu\bar{\nu}}^{NR}=\frac{\sigma_{\nu\bar{\nu}}^{he}}{1+\left(
E/E_{r}\right)  ^{-1}}, \label{SigmaNRnunubar}%
\end{equation}
where $\sigma_{\nu\bar{\nu}}^{he}\simeq8.3\times10^{-4}\mu b$.

We assume that neutrino are non-relativistic even at sufficiently high
redshift, which is a good approximation for $z<10^{2}$ for $m_{\nu}=0.08$
eV/c$^{2}$. Effects of non-zero momentum on the neutrino annihilation
cross-section are studied in \cite{2006APh....25...47D,2013JCAP...08..014L}.
Using the number density of relic neutrinos $n_{C\nu B}\simeq112$ cm$^{-3}$
and the non-resonant cross-section in the high energy limit one can estimate
the horizon for UHE\ neutrinos at highest energies. Solving the Friedmann
equation (see next section) one finds for the redshift $z_{\nu}\simeq84$.

\section{The optical depth and the mean free path}

\label{odandmfp}

In this section we compute the optical depth for the propagation of
UHE\ particles in the Universe. Imposing the condition that it equals unity we
determine the corresponding mean free path. It should be noted that simple
estimates, made in the literature, as well as in previous section, do not take
into account evolution of CMB and C$\nu$B fields with time. The simplest way
to account for cosmic redshift is to compare this estimate of the mean free
path to the expansion scale $c/H_{0}$, where $H_{0}$ is the present day Hubble
parameter, see e.g. \cite{1988A&A...199....1B,2000PhRvD..62i3005S}. In what
follows we describe more rigorous way to take into account both redshift of
particle energy as well as the evolution of CMB and C$\nu$B fields with redshift.

The optical depth along the particle world line $\mathcal{L}$ is defined as
\begin{equation}
\tau=\int_{\mathcal{L}}\sigma j_{\mu}dx^{\mu}, \label{tauWL}%
\end{equation}
where $\sigma$\ is the cross-section of a given process, $j^{\mu}$ is the
4-current of particles, on which the UHE particle scatters, and $dx^{\mu}$ is
the element of the UHE\ particle world line. We assume the Universe is
homogeneous and isotropic, and the background particles are either CMB photons
or C$\nu$B neutrinos. Both have thermal distribution functions, given by%
\begin{equation}
f(\mathcal{E}/kT)=\frac{1}{e^{\left(  \mathcal{E}-\mu\right)  /kT}\pm1},
\label{thermal}%
\end{equation}
where $k$ is the Boltzmann constant, $T$ is the CMB or C$\nu$B temperature,
the sign "$-$" is for photons while the sign "$+$" is for neutrinos,
$\mathcal{E}$ and $\mu$ are the energy and the chemical potential of
background particles (for photons $\mu=0$). Then the optical depth
(\ref{tauWL}) is%
\begin{equation}
\tau(E,t)=\frac{g_{s}}{2\pi^{2}\hbar^{3}c^{3}}\int_{t}^{0}cdt^{\prime}%
\int_{\mathcal{E}_{tr}}^{\infty}\mathcal{E}^{2}d\mathcal{E}f(\mathcal{E}%
)\sigma(E,\mathcal{E},t^{\prime}), \label{tauph}%
\end{equation}
where $\mathcal{E}_{tr}$\ is threshold energy in a given process, $g_{s}=2$ is
the number of helicity states for both protons and neutrinos. Here we assumed
that UHE particles move along light-like geodesics. The integral over time can
be transformed into the integral over redshift by means of the Friedmann
equation. The latter for the flat Universe reads%
\begin{equation}
\left(  \frac{1}{a}\frac{da}{dt}\right)  ^{2}=\frac{8\pi G}{3}\rho,
\label{FREq}%
\end{equation}
where $a$\ is the scale factor, $\rho$ is energy density of the Universe, $G$
is Newton's constant. From this equation, the definition of cosmological
redshift, as well as the definition of the density parameters%
\begin{equation}
a_{0}/a=1+z,\qquad\Omega_{i}=\frac{\rho_{i}}{\rho_{c}},\qquad\rho_{c}%
=\frac{3H_{0}^{2}}{8\pi G}, \label{conventions}%
\end{equation}
where $H_{0}$ and $a_{0}$\ are present time Hubble parameter and scale factor,
respectively, we have%
\begin{equation}
\int_{t}^{0}cdt^{\prime}\longrightarrow\frac{c}{H_{0}}\int_{0}^{z}%
\frac{dz^{\prime}}{\left(  1+z^{\prime}\right)  H(z^{\prime})},
\label{timereshift}%
\end{equation}
where $H_{0}$\ is the Hubble parameter and the function $H(z)$ is given by
\begin{equation}
H(z)=[\Omega_{r}(1+z)^{4}+\Omega_{M}(1+z)^{3}+\Omega_{\Lambda}]^{1/2},
\label{free}%
\end{equation}
and $\Omega_{r}$, $\Omega_{M}$ and $\Omega_{\Lambda}$\ are present densities
of radiation, matter and dark energy, respectively. Then the expression
(\ref{tauph}) can be written as follows%
\begin{equation}
\tau(E,z)=\frac{1}{\pi^{2}\hbar^{3}c^{3}}\frac{c}{H_{0}}\int_{0}^{z}%
\frac{dz^{\prime}}{\left(  1+z^{\prime}\right)  H(z^{\prime})}\int%
_{\mathcal{E}_{tr}}^{\infty}\mathcal{E}^{2}d\mathcal{E}f(\mathcal{E}%
)\sigma(E,\mathcal{E},z^{\prime}). \label{opacity00}%
\end{equation}
Cosmic expansion results in the energy and temperature dependence on redshift%
\begin{equation}
T=(1+z)T_{0},\quad\mathcal{E}=(1+z)\mathcal{E}_{0},\quad E=(1+z)E_{0},
\label{rede}%
\end{equation}
where temperature $T_{0,\gamma}\simeq2.725$ K for photons, $T_{0,\nu}=\left(
4/11\right)  ^{1/3}\simeq1.95$ K for neutrinos and energies $E_{0}%
,\mathcal{E}_{0}$ are measured at the present time.

The second integral in (\ref{opacity00}) can be simplified, provided two
conditions are fulfilled: a) the cross-section does not depend on the energy
of background particle and b)\ there is no threshold in the given process
($\mathcal{E}_{tr}=0$). In this case one has%
\begin{equation}
\frac{1}{\pi^{2}\hbar^{3}c^{3}}\int_{0}^{\infty}\mathcal{E}^{2}d\mathcal{E}%
f(\mathcal{E})\sigma(E,z)=\sigma(E,z)n(z)=\sigma(E,z)n_{0}\left(  1+z\right)
^{3}, \label{secondint}%
\end{equation}
where $n_{0}$ is present number density and it stands for either
\[
n_{0,\gamma}\approx\frac{2\zeta\left(  3\right)  }{\pi^{2}}\left(  \frac
{\hbar}{m\,c}\right)  ^{-3}\left(  \frac{kT_{0}}{m_{e}c^{2}}\right)
^{3}\simeq411\ \text{cm}^{-3}%
\]
for photons, or $n_{0,\nu}=3/4\left(  T_{0,\nu}/T_{0,\gamma}\right)
^{3}\simeq113$\ cm$^{-3}$\ for neutrinos. Then eq. (\ref{tauph}) becomes%
\begin{equation}
\tau(E,z)=n_{0}\frac{c}{H_{0}}\int_{0}^{z}\frac{\sigma(E,z^{\prime})\left(
1+z^{\prime}\right)  ^{2}dz^{\prime}}{H(z^{\prime})}. \label{taun}%
\end{equation}
When the cross-section is just a constant, the integral (\ref{taun}) can be
readily performed. Assuming $\Omega_{r}\simeq9.2\times10^{-5}$, $\Omega
_{M}\simeq0.315$, $\Omega_{\Lambda}\simeq0.685$ and $H_{0}=67.3$ km/s/Mpc
\cite{2014A&A...571A..16P} in the matter dominated epoch we have%
\[
\int_{0}^{z}\frac{\left(  1+z^{\prime}\right)  ^{2}dz^{\prime}}{[\Omega
_{M}(1+z^{\prime})^{3}+\Omega_{\Lambda}]^{1/2}}\simeq\left\{
\begin{array}
[c]{cc}%
1.045z, & z\ll1,\\
1.006z^{3/2}, & z\gg1.
\end{array}
\right.
\]

The mean free path is defined by the condition $\tau(E_{0},z)=1$. For the
constant cross-section $\sigma$ at low redshift $z\ll1$ we get the traditional
definition $\lambda=\left(  \sigma n\right)  ^{-1}$\ used above. For high
redshift $z\gg1$ one can define the redshift, corresponding to the mean free
path as
\begin{equation}
z_{\lambda}=\left(  \frac{n_{0}\sigma c}{H_{0}}\right)  ^{-2/3}\simeq
8.9\left(  \frac{n_{0}}{n_{0,\gamma}}\frac{\sigma}{10^{-8}\sigma_{T}}\right)
^{-2/3}. \label{zlambda}%
\end{equation}
Using this equation we obtain for UHE\ neutrinos with highest energies
$z_{\lambda}\simeq84$.

\section{The mean energy loss distance}

\label{meld}

When UHE\ particle annihilates in a given process, such as in the case of
Breit-Wheeler one (\ref{BWprocess}), the mean free path correpsonds to the
horizon defined above.

Another possibility is that the particle is not annihilated in a given
process, but scattered, such as in the case of proton producing the pion
(\ref{photopion}). When the energy loss in single scattering corresponds to a
large fraction of UHE particle energy, the situation is similar to the case of
annihilation. However, UHE particle may lose only a small fraction of its
energy, as in the case of Bethe-Heitler process (\ref{ppair}). Here another
relevant quantity corresponds the particle horizon defined above is the mean
energy loss distance $\tilde{\lambda}$. We define it following
\cite{1970PhRvD...1.1596B} as%
\begin{equation}
\tilde{\lambda}^{-1}=\left(  \frac{1}{E}\frac{dE}{cdt}\right)  . \label{melp}%
\end{equation}
Then we evaluate the quantity%
\begin{equation}
\tilde{\tau}=\int_{t}^{0}\frac{cdt}{\tilde{\lambda}}=\frac{c}{H_{0}}\int%
_{0}^{z}\frac{dz^{\prime}}{\tilde{\lambda}\left(  1+z^{\prime}\right)
H(z^{\prime})}. \label{tautilde}%
\end{equation}
It is computed below for the Bethe-Heitler process.

\section{Results}

\label{results}

Now we apply the method developed in the previous section to the computation
of the mean free path for UHE photons, protons and neutrinos, as well as the
mean energy loss distance for protons interacting via the Bethe-Heitler process.

\subsection{Photons}

First, we consider cosmic limits on propagation of UHE\ photons. In the
Breit-Wheeler process (\ref{BWprocess}) the cross-section depends on both
energies through the definition (\ref{substitute}). When one considers all
possible orientations of CMB photons additional averaging over their angular
distribution has to be performed \cite{Nikishov1961,1967PhRv..155.1408G}. The
resulting averaged cross section differs from eq. (\ref{BW section0}). The
useful approximations for this quanity can be found e.g. in
\cite{1967PhRv..155.1408G,1983Afz....19..323A,1990MNRAS.245..453C}. We use the
accurate expression given by eq. (3.23) in Ref. \cite{Aharonian2003}:%
\begin{align}
\bar{\sigma}_{\gamma\gamma}\left(  x\right)   &  =\frac{3}{2}\sigma_{T}%
\Sigma\left(  x\right)  ,\qquad\Sigma\left(  x\right)  =\frac{1}{x^{2}}\left[
\left(  x+\frac{1}{2}\log x-\frac{1}{6}+\frac{1}{2x}\right)  \times\right.
\label{avBW}\\
&  \left.  \times\log\left(  \sqrt{x}+\sqrt{x-1}\right)  -\left(  x+\frac
{4}{9}-\frac{1}{9x}\right)  \sqrt{1-\frac{1}{x}}\right]  .\nonumber
\end{align}
We change our variables in eq. (\ref{opacity00}) and get%
\begin{equation}
\tau_{\gamma\gamma}(E_{0},z)=\frac{A}{y_{0}^{3}}\int_{0}^{z}\frac
{1}{(1+z^{\prime})^{4}}\frac{dz^{\prime}}{H\left(  z^{\prime}\right)  }%
\int_{1}^{\infty}\frac{x^{2}dx}{\exp(x/y)-1}\Sigma\left(  x\right)
,\label{opa1}%
\end{equation}
where%
\begin{equation}
A=\frac{4\alpha^{2}}{\pi}\frac{c}{H_{0}}\left(  \frac{\hbar}{m\,c}\right)
^{-1}\left(  \frac{kT_{0}}{m_{e}c^{2}}\right)  ^{3}\approx2.37\times10^{6},
\end{equation}
and
\begin{equation}
y=y_{0}(1+z)^{2};\quad y_{0}=\frac{E_{0}}{m_{e}c^{2}}\frac{kT_{0}}{m_{e}c^{2}%
},\label{dete}%
\end{equation}
and $y_{0}$ is the energy $E_{0}$ in units of the critical energy
$E_{BW}=\left(  m_{e}c^{2}\right)  ^{2}/kT_{0}\simeq1.11\times10^{15}$ eV. The
intergal over energy can be evaluated numerically and we find a reasonable fit%
\begin{equation}
F_{1}\left(  y\right)  =0.839\left(  y^{2.1}+2\times10^{-8}y^{2.8}\right)
\exp\left(  -\frac{1.1}{y}\right)  .\label{F1}%
\end{equation}
Then eq. (\ref{opa1}) becomes%
\begin{equation}
\tau_{\gamma\gamma}(y_{0},z)=\frac{A}{y_{0}^{3}}\int_{0}^{z}\frac
{1}{(1+z^{\prime})^{4}}\frac{dz^{\prime}}{H\left(  z^{\prime}\right)  }%
F_{1}\left[  y_{0}(1+z)^{2}\right]  .\label{opa1F}%
\end{equation}
This integral is also evaluated numerically.

In the low energy $E\ll E_{BW}$\ and high redshift $z\gg1$\ limit the integral
(\ref{opa1F}) can be evaluated analytically. The redshift corresponding to the
mean free path in this limit is
\begin{equation}
z_{\lambda,BW}\simeq0.21\left(  \frac{E}{E_{BW}}\right)  ^{-1/2}. \label{zbw}%
\end{equation}
This result is known as the Fazio-Stecker relation \cite{1970Natur.226..135F}, see their eq. (9).

In addition to the Breit-Wheeler process (\ref{BWprocess}), following
\cite{1997ApJ...487L...9C} we consider also the double pair production process
(\ref{dpp}) with the cross-section defined in (\ref{sigmaep2}). This process
is relevant for the highest energies. The optical depth for this process is%
\begin{align}
\tau_{\gamma\gamma}^{dpp}(y_{0},z)  &  =\frac{B}{y_{0}^{3}}\int_{0}^{z}%
\frac{1}{(1+z^{\prime})^{4}}\frac{dz^{\prime}}{H\left(  z^{\prime}\right)
}\int_{2}^{\infty}\sigma_{dpp}\frac{x^{2}dx}{\exp(x/y)-1}=\label{taud}\\
&  =\frac{B}{y_{0}^{3}}\int_{0}^{z}\frac{1}{(1+z^{\prime})^{4}}\frac
{dz^{\prime}}{H\left(  z^{\prime}\right)  }F_{2}\left[  y_{0}(1+z^{\prime
})^{2}\right]  ,\nonumber
\end{align}
where%
\begin{align}
F_{2}\left(  y\right)   &  =\frac{8}{3}-4i\pi y-4y\log\left[  \exp\left(
\frac{2}{y}\right)  -1\right]  -\\
&  -4y^{2}\mathrm{PolyLog}\left[  2,\exp\left(  \frac{2}{y}\right)  \right]
+2y^{3}\mathrm{PolyLog}\left[  3,\exp\left(  \frac{2}{y}\right)  \right]
.\nonumber
\end{align}
and $B\approx15.3$.

\begin{figure}[pth]
\centering
\includegraphics[width=0.8\hsize]{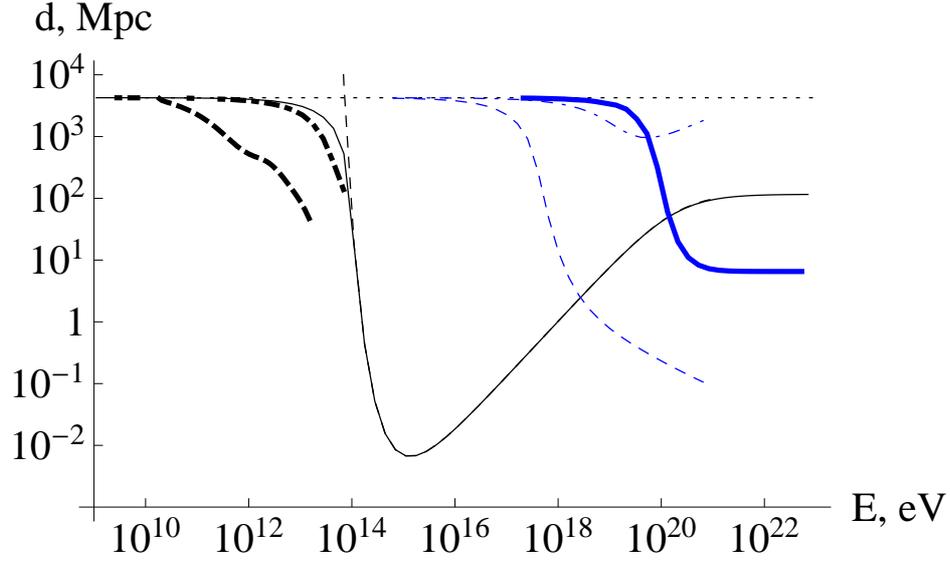} \caption{The mean free path, measured in
megaparsecs as a function of energy $E$ of UHE particles, measured in
electronvolts. In the region above the curves the optical depth is larger than
unity. Thin black curve shows the mean free path of UHE photons. Dashed black curve
shows the photon mean free path computed without accounting for cosmological
evolution (imposing $z=0$). Thick blue dashed curve shows the boundary of
transparency for extragalactic background light (EBL), according to the baseline model of Inoue et al.
\cite{2013ApJ...768..197I}. Thick blue curve shows the mean free path of UHE
protons (GZK limit). Blue dashed (dotted-dashed) curve shows the mean free
path (mean energy loss distance) for UHE protons due to Bethe-Heitler process.
Dotted horizontal line shows cosmological horizon.}%
\label{d}%
\end{figure}
\begin{figure}[ptb]
\centering
\includegraphics[width=0.8\hsize]{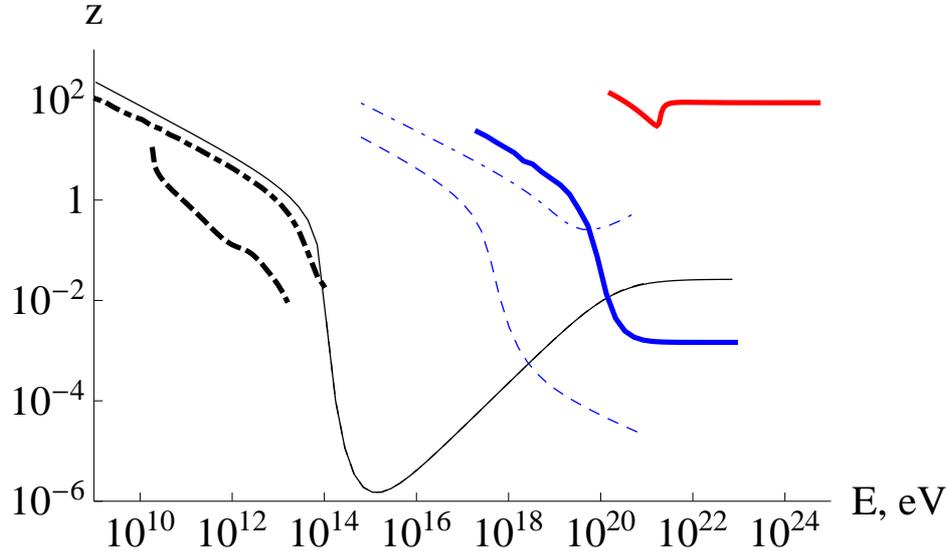} \caption{The same as in Fig. \ref{d} for
the distance measured in cosmological redshift. Red thick curve shows the
mean free path of UHE neutrinos.}%
\label{z}%
\end{figure}

The condition $\tau(y_{0},z)=\tau_{\gamma\gamma}(y_{0}%
,z)+\tau_{\gamma\gamma}^{dpp}(y_{0},z)=1$ in eqs.\ (\ref{opa1}) and
(\ref{taud}) determines the mean free path of UHE photons. This mean free path
is shown in Fig. \ref{d} in megaparsecs and in Fig. \ref{z} in cosmological
redshift by the solid curve. 
The region above the solid curve is opaque for high energy
photons. In addition, thick black dashed line shows the boundary of transparency for
EBL, according to the baseline model of Inoue et al. \cite{2013ApJ...768..197I}, while dotted line in Fig.
\ref{d} shows the cosmological horizon $z=\infty$.

For the distance smaller than a critical value of about $d_{c}=6.8$ kpc, the
CMB is transparent to high-energy photons with arbitrary energy. For larger
distances there are two branches of solutions for the condition $\tau
_{\gamma\gamma}(y_{0},z)=1$, respectively corresponding to the different
energy-dependence of the average cross-section (\ref{avBW}). This average
cross-section $\bar{\sigma}_{\gamma\gamma}\left(  x\right)  $ increases with
the center of mass energy $x$ from the energy threshold $x=1$ to $x\simeq3.5$,
and decreases from $x\simeq3.5$ to $x\rightarrow\infty$. The energy of the UHE
photon corresponding to the critical distance $d_{c}$ is about $1.11$ PeV,
which separates two branches of the solution. The double pair production
process (\ref{dpp}) is relevant for the highest energies, as expected. Photons
with energies above $10$ PeV are absorbed by the double pair production if
they are emitted at redshift above $z\simeq0.03$ (distance about 120 Mpc).

For comparison we show by the dashed curve also the mean free path computed at
$z=0$, namely neglecting cosmological expansion, see e.g.
\cite{1997ApJ...487L...9C}.

We also show by the dashdotted thick curve the mean free path for the
photon-photon scattering which follows from the Euler-Heisenberg lagrangian,
see e.g. \cite{1982els..book.....B,2010PhR...487....1R}. We will discuss this
process in a separate publication \cite{Batebi2015}.

Finally, the black dotted thick curve shows the horizon of photons with energies above 20 GeV and below 100 TeV, which is determined by their interaction with the EBL. The latest EBL model \cite{{2013ApJ...768..197I}} is used. It is clear, that the contribution of CMB photons gives the absolute upper limit on the mean free path. In the energy range between 1 GeV and 20 GeV the propagation of high energy photons is limited only by the CMB radiation.

\subsection{Protons}

Second, we consider the propagation of UHE protons, accelerated in a source
located at a cosmological distance from Earth. First, considering the
photopion process (\ref{photopion}) we use the method developed in the
previous section and compute the GZK limit
\cite{1966PhRvL..16..748G,1966JETPL...4...78Z}. This limit applies to protons
and other charged particles, leading to the existence of a cutoff in the
observed spectrum of (UHE) cosmic rays at about $10^{20}$ eV. For the
photopion process (\ref{photopion}) one can use the simple expression
(\ref{taun}) with the constant cross-section (\ref{sigmapgamma}). However, we
compute the optical depth in the same way as in the case of double pair
production, using eq. (\ref{taud}) with different value of the constant
$B^{\prime}\approx253$.

The mean free path due to photopion process is shown by the blue thick curve in
Fig. \ref{d} in megaparsecs and in Fig. \ref{z} in cosmological redshift. From
Fig. \ref{d} it appears that for energies well below $E_{p\gamma}%
\simeq3.3\times10^{20}$ eV the GZK limit approaches the cosmological horizon.
Instead, from Fig. \ref{z} it follows that the mean free path measured in
redshift, below the energy $E_{p\gamma}$, increases with decreasing energy as
a power law, which is a consequence of eq. (\ref{zlambda}).

Similarly to the Breit-Wheeler case, the integral (\ref{taud}) is evaluated
analytically in the low energy $E\ll E_{p\gamma}$\ and high redshift $z\gg
1$\ limit, with the result%
\begin{equation}
z_{\lambda,GZK}\simeq0.57\left(  \frac{E}{E_{p\gamma}}\right)  ^{-1/2}.
\label{zgzk}%
\end{equation}

We also evaluate the mean free path due to the Bethe-Heitler process
(\ref{ppair}). Since cross-sections (\ref{sigmaphotopairthr}) and
(\ref{sigmaphotopair})\ are given in the proton rest frame, one has to
transform photon energy to this reference frame using%
\begin{equation}
\mathcal{E}^{\prime}=2\Gamma\mathcal{E=}2\frac{E\mathcal{E}}{m_{p}c^{2}},
\label{prflabtrans}%
\end{equation}
where the primed quantity corresponds to the proton rest frame, while unprimed
quantities to laboratory reference frame.\ Then it is convenient to make use
of the same type of variable change as before for the Breit-Wheeler
process\footnote{We assume that UHE\ protons collide with the with
CMB\ photons head on. More accurate calculation with average over angular
distribution of the CMB\ photons does not change qualitatively our results.},
with a difference that instead of electron mass squared a product of electron
and proton masses arises, namely%
\begin{equation}
\bar{x}=2\frac{E\mathcal{E}}{m_{e}m_{p}c^{4}},\qquad\bar{y}_{0}=2\frac{E_{0}%
}{m_{p}c^{2}}\frac{kT_{0}}{m_{e}c^{2}}=\frac{E_{0}}{E_{BH}}.
\label{pgammavars}%
\end{equation}
The optical depth is computed in the laboratory frame as follows\footnote{Note
that the integral over energy is not transformed to the proton rest frame, as
done e.g. in \cite{1970PhRvD...1.1596B}. Instead, only a change of variables
is performed in the integral (\ref{BHod}).}%
\begin{align}
\tau_{p\gamma}(\bar{y}_{0},z)  &  =\frac{1}{\pi^{2}}\frac{c}{H_{0}}\left(
\frac{\hbar}{m\,c}\right)  ^{-3}\left(  \frac{kT_{0}}{m_{e}c^{2}}\right)
^{3}\times\label{BHod}\\
&  \times\frac{1}{\bar{y}_{0}^{3}}\int_{0}^{z}\frac{1}{(1+z^{\prime})^{4}%
}\frac{dz^{\prime}}{H\left(  z^{\prime}\right)  }\int_{2}^{\infty}\frac
{\bar{x}^{2}d\bar{x}}{\exp(\bar{x}/\bar{y})-1}\sigma_{BH}(x).\nonumber
\end{align}
The intergal over energy can be evaluated numerically and we find a reasonable
fit%
\begin{equation}
F_{3}\left(  \bar{y}\right)  =\frac{86.15\exp\left(  -\frac{2}{\bar{y}%
}\right)  }{10^{3\bar{y}-3.47}+\bar{y}^{-3}}. \label{opa3}%
\end{equation}
Then eq. (\ref{opa1}) becomes%
\begin{equation}
\tau_{p\gamma}(\bar{y}_{0},z)=\frac{C}{\bar{y}_{0}^{3}}\int_{0}^{z}\frac
{1}{(1+z^{\prime})}\frac{dz^{\prime}}{H\left(  z^{\prime}\right)  }%
F_{3}\left[  \bar{y}_{0}(1+z)^{2}\right]  , \label{opa3F}%
\end{equation}
where%
\begin{equation}
C=\frac{2\alpha^{3}}{3\pi}\frac{c}{H_{0}}\left(  \frac{\hbar}{m\,c}\right)
^{-1}\left(  \frac{kT_{0}}{m_{e}c^{2}}\right)  ^{3}\approx2863.
\end{equation}
The mean free path for protons interacting via the Bethe-Heitler process is
shown by blue dashed curve in Fig. \ref{d} in megaparsecs and in Fig. \ref{z}
in cosmological redshift. The integral (\ref{opa3F}) is evaluated analytically
in the low energy $E\ll E_{BH}$\ and high redshift $z\gg1$\ limit, with the
result%
\begin{equation}
z_{\lambda,BH}\simeq0.43\left(  \frac{E}{E_{BH}}\right)  ^{-1/2}. \label{zbh}%
\end{equation}

As discussed before, at energies above $10^{18}$ eV the mean free path is
relatively small, about a few hundred kiloparsecs, and it quickly decreases
with increasing energy. This is in contrast with the large mean energy loss
path, which is above $1$ Gpc at energies $10^{18}-10^{20}$ eV, see e.g.
\cite{1988A&A...199....1B,2000PhRvD..62i3005S}. It means, that before UHE
proton starts to lose its energy, it is scattered many hundred times
\cite{2006NJPh....8..122D}. At each of this scattering the proton recoils,
being deflected by a small angle measured in the laboratory reference frame.
From the analysis of the cross-section as the function of recoil angle
\cite{1950PhRv...80..189J}, see also \cite{1969RvMP...41..581M} in the proton
rest frame, it follows that in the high energy limit the photon recoils in the
plane, orthogonal to the incident photon. It implies that each scattering
produces a deflection of the UHE\ proton in the laboratory frame by angle
$\sim1/\gamma$, where $\gamma=E/\left(  m_{p}c^{2}\right)  $\ is proton
Lorentz factor. The number of scatterings is given approximately by $\tau$.

One can compute the mean energy loss distance defined in (\ref{melp}) and then
evaluate the quantity (\ref{tautilde}) using eq. (18)-(20) in
\cite{1970PhRvD...1.1596B} to obtain%
\begin{equation}
\tilde{\tau}=\frac{D}{y_{0}^{3}}\int_{0}^{z}\frac{dz^{\prime}}{\left(
1+z^{\prime}\right)  ^{4}H(z^{\prime})}\int_{2}^{\infty}\frac{d\bar{x}}%
{\exp(\frac{\bar{x}}{\bar{y}})-1}\phi\left(  \bar{x}\right)  ,
\label{tautildep}%
\end{equation}
where the function $\phi\left(  \bar{x}\right)  $\ is given in eq. (16) in
\cite{1970PhRvD...1.1596B}
\begin{equation}
\phi\left(  \bar{x}\right)  =\bar{x}\left[  -86.07+50.95\log\bar
{x}-14.45\left(  \log\bar{x}\right)  ^{2}+2.667\left(  \log\bar{x}\right)
^{3}\right]  \label{phif}%
\end{equation}
and
\begin{equation}
D=\frac{2}{\pi^{2}}\alpha^{3}\frac{m_{e}}{m_{p}}\frac{c}{H_{0}}\left(
\frac{\hbar}{m\,c}\right)  ^{-1}\left(  \frac{kT_{0}}{m_{e}c^{2}}\right)
^{3}.
\end{equation}
From the condition $\tilde{\tau}=1$ we determine the mean energy loss distance
$\tilde{\lambda}$. This distance is shown by blue dash-dotted curve in\ Fig.
\ref{d} in megaparsecs and in Fig. \ref{z} in cosmological redshift.

\begin{figure}[ptb]
\centering
\includegraphics[width=0.7\hsize]{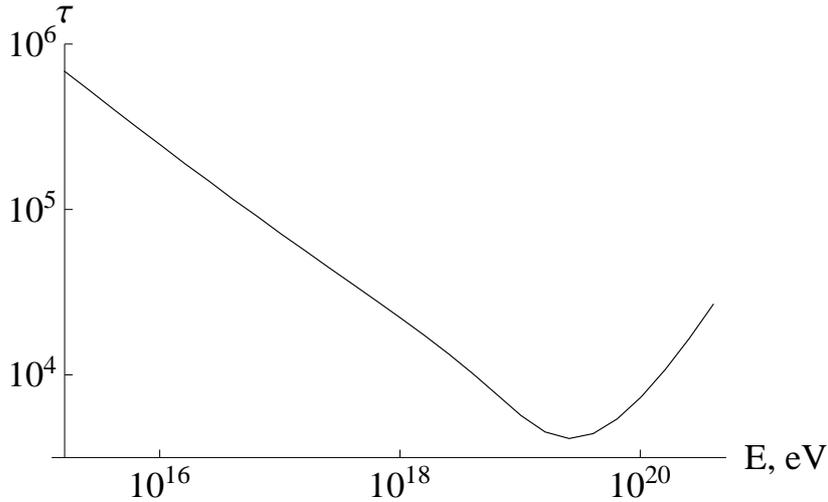} \caption{The optical depth at the
mean energy distance for the Bethe-Heitler process.}%
\label{tauBH}%
\end{figure}
\begin{figure}[ptb]
\centering
\includegraphics[width=0.7\hsize]{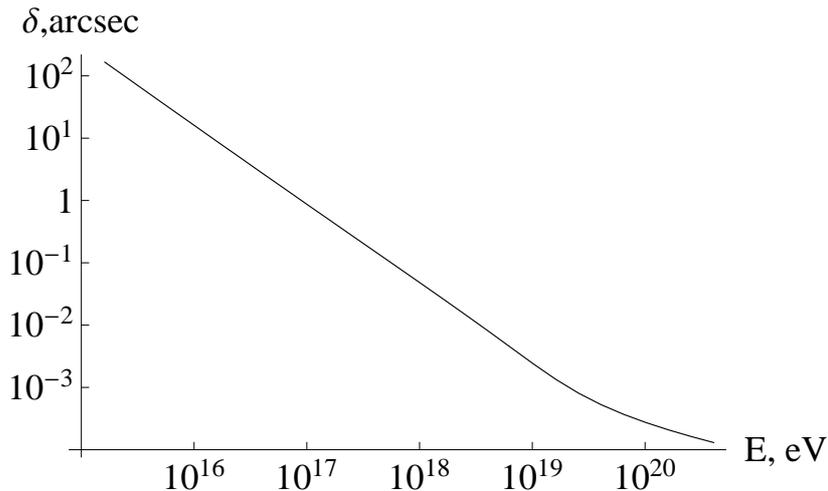} \caption{The average deflection angle
of UHE protons as function of proton energy for sources located at the mean
energy loss distance for the Bethe-Heitler process.}%
\label{delta}%
\end{figure}

We evaluate the optical depth (\ref{BHod}) at the redshift,
corresponding to $\tilde{\lambda}$ for energies in the range between $10^{15}$
and $10^{20}$\ eV and find it in the range$\ 10^{4}$ to $10^{5}$, see Fig.
\ref{tauBH}. Since the deflection at each interaction is small, the average
number of interactions is proportional to the optical depth. The average
deflection angle is then%
\begin{equation}
\delta\sim\frac{\sqrt{\tau\left(  E\right)  }}{\gamma}. \label{defangle}%
\end{equation}
We find an average\ deflection angle of UHE\ protons as a function of proton
energy for sources located at the mean energy loss distance and show it in
Fig. \ref{delta}. At energies $E=10^{16}$eV, this angle is about $\delta
\sim15^{\prime\prime}$\ and it decreases down to $\delta\sim2.4$\ mas for
$E=10^{19}$eV. This latter scale is larger than the angular size of distant
blazars \cite{2008JPhCS.131a2054G}. Such deflection, although small compared
to deflection in galactic magnetic field \cite{1998ApJ...492..200M},
inevitably leads to dimming of point sources of UHE\ protons. It makes also
more difficult to detect distant sources. 

\subsection{Neutrinos}

Third, we consider the propagation of UHE neutrinos. Such neutrinos can be
produced in the source of UHE\ cosmic rays in decay of secondary pions
$\pi^{+}\longrightarrow\mu^{+}+\nu_{\mu}$ or secondary neutrons
$n\longrightarrow p+e^{-}+\bar{\nu}_{e}$ \cite{2006NJPh....8..122D}.
UHE\ neutrino can be produced also in some extensions of the standard model of
particle physics \cite{2006JPhCS..39..393R}. Such UHE\ neutrino interacts with
the C$\nu$B via the process (\ref{Rnunubar}).

The cross-section of this process has a resonance, and it approaches a
constant for highest energies. We compute the optical depth, which instead of
eq. (\ref{opacity00}) is given by%
\begin{align}
\tau_{\nu\nu}(E,z)  &  =\frac{1}{\pi^{2}\hbar^{3}c^{3}}\frac{c}{H_{0}}\int%
_{0}^{z}\frac{dz^{\prime}}{\left(  1+z^{\prime}\right)  H(z^{\prime})}%
\int_{\mathcal{E}_{tr}}^{\infty}\mathcal{E}\sqrt{\mathcal{E}^{2}-\left(
m_{\nu}c^{2}\right)  ^{2}}d\mathcal{E}f(\mathcal{E})\sigma(E,z)\approx
\label{taunu}\\
&  \frac{1}{\pi^{2}\hbar^{3}c^{3}}\frac{c}{H_{0}}n_{0,\nu}\int_{0}^{z}%
\frac{\left(  1+z^{\prime}\right)  ^{2}dz^{\prime}}{H(z^{\prime})}\sigma(E(1+z\prime)),\nonumber
\end{align}
using the cross-sections given in the laboratory reference frame by eqs.
(\ref{SigmaRnunubar}) and (\ref{SigmaNRnunubar}). The mean free path for
neutrinos, measured in cosmological redshift, is shown in Fig. \ref{z} by the
thick red curve. Since the characteristic redshifts are high, this curve
practically coincides with the horizon, when measured in megaparsecs, so we do
not show it in Fig. \ref{d}. It is clear that the Breit-Wigner resonance in
the cross-section decreases the mean free path in a wide range of energies. The lowest redshift for
$E\simeq E_{r}$ at which the Universe is transparent for UHE\ neutrinos is
$z_{min}\simeq30$. The resonance produces a dip around $E_{r}/(1+z_{min})\simeq1.7\times10^{21}\left(  m_{\nu}/0.08\text{ eV}\right)^{-1}$ eV, where $z_{min}\simeq 30$. Additional broadening of the resonance, due to thermal effect, is
discussed in detail in \cite{2013JCAP...08..014L}. At higher energies the corresponding redshift is $z\simeq87$.

Similarly to the previous cases, in the low energy $E\ll E_{p\gamma}$\ and high redshift $z\gg
1$\ limit, we find
\begin{equation}
z_{\lambda,\nu}\simeq14\left(  \frac{E}{E_{r}}\right)  ^{-2/5}.
\label{znu}%
\end{equation}

\section{Conclusions}

We reviewed cosmic limits on propagation of ultra high energy particles such
as photons, protons and neutrinos, set up by their interactions with the
cosmic background of photons and neutrinos. In doing so we take into account
explicitly cosmic evolution of both cosmic backgrounds, and redshift of UHE
particle energy. This is in contrast with majority of the literature, where
corresponding mean free paths are found at present epoch, neglecting cosmic
expansion. A number of new results were obtained, in particular:

\begin{itemize}
\item for UHE\ photons the contribution of CMB photons gives the absolute
upper limit on the mean free path. At high redshift, where other radiation
backgrounds, such as EBL are absent, the CMB radiation limits the propagation
of UHE\ photons at energies above GeV.

\item for UHE\ protons the mean free path due to Bethe-Heitler process appears
to be much shorter than the mean energy loss distance. This results in
multiple deflections suffered by UHE\ protons, before they start to lose
energy in the energy range $10^{16}-10^{20}$ eV. Such deflections result in
dimming of point sources of UHE\ protons, which makes it more difficult to
detect them.

\item for UHE\ neutrinos for the first time we compute the horizon as a
function of redshift. We found that the Universe is transparent of
UHE\ neutrinos at redshifts $z<30$, near the Breit-Wigner resonance at
$E_{r}\simeq5.2\times10^{22}\left(  m_{\nu}/0.08\text{ eV}\right)  ^{-1}$ eV,
and it is transparent at redshifts $z<87$ at higher energies.

\item Remarkably, in the low energy and high redshift limit, the Fazio-Stecker relation \cite{1970Natur.226..135F} holds for all processes with exception of neutrinos, and it is given by a universal expression $z_{\lambda
}\simeq\mathcal{O}\left(  1\right)  \left(  \frac{E}{E_{thr}}\right)  ^{-1/2}$,
where $E_{thr}$\ is the characteristic (e.g. threshold) energy for a given process. In the case of neutrinos similar power law exists $z_{\lambda}\propto E^{-2/5}$.
\end{itemize}



\end{document}